\documentclass[journal]{IEEEtran}

\usepackage{cite}
\usepackage{amsmath,amssymb,amsfonts}
\usepackage{algorithmic}
\usepackage{graphicx}
\usepackage{textcomp}
\usepackage{xcolor}
\usepackage{footnote}
\usepackage{hyperref}
\usepackage{xcolor}
\usepackage{tabu}
\usepackage{makecell}
\usepackage{float}
\usepackage{adjustbox}
\usepackage{multirow,booktabs}
\usepackage[inline]{enumitem}
\usepackage{lipsum}
\usepackage[center]{caption}
\usepackage{subcaption}
\usepackage{soul}
\usepackage{flushend}
\usepackage{comment}
\begin{document}

\title{IEEE 802.11be Multilink Operation}
\title{Multi-link Operation in IEEE 802.11be WLANs}

\author{Álvaro~López-Raventós, and Boris~Bellalta
\thanks{All the authors are with the \textit{Wireless Networking (WN)} research group at \textit{Universitat Pompeu Fabra, Barcelona, Spain} (e-mail: {alvaro.lopez@upf.edu, boris.bellalta}@upf.edu). This  work  has  been  partially  supported by the Spanish Government under grant WINDMAL PGC2018-099959-B-I00 (MCIU/AEI/FEDER,UE), by the Catalan Government under grant 2017-SGR-1188, and Cisco.}}
\maketitle

\begin{abstract}
The multi-link operation (MLO) is a new feature proposed to be part of the IEEE 802.11be Extremely High Throughput (EHT) amendment. Such feature represents a paradigm shift towards multi-link communications, as nodes will be allowed to transmit and receive data over multiple radio interfaces concurrently. To make it possible, the 802.11be Task Group has proposed different modifications in regards to nodes' architecture, transmission operation, and management functionalities. This article reviews such changes and tackles the question of how traffic should be distributed over multiple links, as it is still unresolved. To that end, we evaluate different load balancing strategies over the active links. Results show that in high load, dense and complex scenarios, implementing congestion aware load balancing policies to significantly enhance next-generation WLAN performance using MLO is a must.
\end{abstract}

\begin{IEEEkeywords}
IEEE 802.11be, multi-link operation, traffic allocation, WLANs.
\end{IEEEkeywords}


\section{Introduction}

Commonly known as WiFi, the IEEE 802.11 standard was released back in the late 90s, with the aim to provide a low complex, and cost efficient, wireless connectivity solution. Currently in its 6th generation, the proliferation of WiFi has been driven by the constant revision of the standard, since periodic amendments have made possible to face the increasing requirements of newer use-cases. Wireless data services will continue to grow, with upcoming applications, such as virtual/augmented reality, video/game streaming and cloud based services, requesting vasts amounts of data with the most demanding throughput, latency, and reliability requirements. To address such expectations, the 802.11be Task Group (TGbe) was created in May 2019 to address the development of new specifications to fuel the upcoming WiFi~7.

Referred to as IEEE 802.11be Extremely High Throughput (EHT)~\cite{draft11be}, this amendment aims to increase the WiFi throughput, while reducing the end-to-end latency and improving the reliability of communications~\cite{lopez2019ieee}. For such purpose, the Multi-link Operation (MLO\footnote{Throughout this paper, we will refer to the multi-band/multi-channel operation feature as the MLO, following the notation of the TGbe.}) is considered a main candidate feature, as it promotes the use of multiple wireless interfaces to allow concurrent data transmission and reception in access points (APs) and stations (STAs) with dual- or tri-band capabilities. 

Indeed, the interest in the use of the MLO framework is rapidly increasing. Latency in real-time applications has been already studied in~\cite{naik2021can,lacalle2021analysis,carrascosa2021experimental}, showing that MLO is able to significantly reduce worst-case latency. Besides, authors in~\cite{lacalle2021analysis} extended their analysis to evaluate the reliability over multiple links, showing a high delivery rate when having multiple uncorrelated links. Also, an end time alignment mechanism to allow the use of parallel downlink transmissions through different links to stations without simultaneous transmit and receive (STR) capability is presented in~\cite{naribole2020simul}. Such approach is intended to maximize the spectrum efficiency. Analogously, an opportunistic backoff mechanism is proposed in~\cite{naribole2020simultaneous} to allow non-STR stations to resume their backoff timers, if an ongoing transmission is identified to not cause a collision. Authors in~\cite{yang2019ap} suggest that the use MLO per se may not be sufficient enough without coordination between APs, proposing a coordination framework to achieve high throughput requirements in high density areas. 

The integration of a framework capable to operate at the same time over multiple wireless interfaces brings up new challenges and research opportunities. In this context, we find that MLO compliant devices will have the ability to transmit and receive packets with different quality-of-service (QoS) requirements over multiple links. Such functionality, which was not allowed in past amendments, is called traffic identifier (TID) to link mapping, and opens up to conceive new traffic management mechanisms. For instance, we may find all TIDs to be assigned to all links, allowing a full adaptive load balancing strategy, as traffic may be moved partially or fully between multiple links. On the contrary, other approaches may rely on having a dedicated link assigned to an specific QoS traffic, which implies a more rigid and less flexible load balancing solution, but ensuring that only traffic with the same QoS requirements share the same set of resources.

\begin{figure*}[ht]
    \centering
    \includegraphics[width=\linewidth]{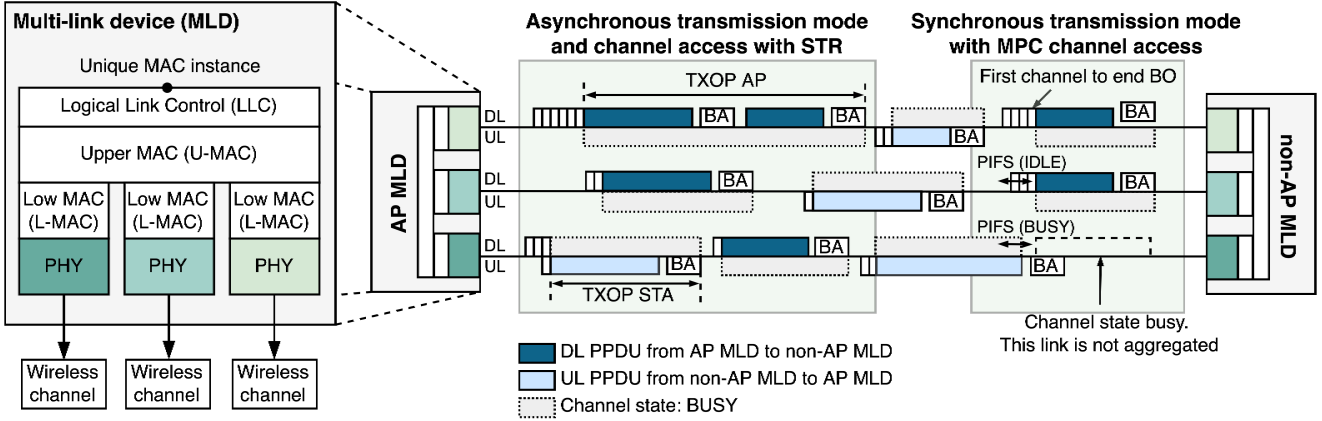}
    \caption{Multi-link architecture and transmission modes representation. Each PHY color represents a different band/channel for each one of the different interfaces.}
    \label{fig:FigOverview}
\end{figure*}

In this article, we assess different allocation strategies that follow either an adaptive or link-dedicated implementation, with the aim to provide some insights on how to distribute the traffic across multiple interfaces, and how it may affect to the network performance. The main contributions are:
\begin{itemize}
    \item We provide a comprehensive overview on how the MLO framework is being devised by the TGbe, pointing out the different modifications in regards the nodes' architectural changes, transmission modes and management functions.
    \item We discuss the potential benefits, and challenging issues related to the overviewed modifications. Also, we point out some open issues and research directions faced by the MLO framework.
    \item We assess different policy-based strategies in order to tackle the traffic allocation problem. Also, we adopt the TID-to-link mapping functionality to showcase its implementation, benefits and drawbacks.
    \item We evaluate the presented strategies under different traffic requirements, showing that a link-dedicated approach may not be suitable neither in high dense, nor high load use-case scenarios.
\end{itemize}


\section{Multi-link operation}

In the following, we explore the different, and the most relevant, proposals that are likely to be included in the IEEE 802.11be amendment regarding the MLO feature.

\subsection{Architecture}\label{sub:architecture}

The first architectural change is found in the redefinition of classical APs or STAs into the so-called multi-link capable devices (MLDs). Either AP~MLDs or STA~MLDs, refer to single devices with multiple wireless interfaces\footnote{Instead of interfaces, the TGbe defines them as affiliated AP/STAs. However, for sake of simplicity, and comprehensive purposes, we will keep referring to them as interfaces.}. The most relevant aspect about that remains on the fact that MLDs will provide a unique MAC instance to the upper layers, without losing the independent parameters of each interface. To achieve that, TGbe proposed to divide the MAC sub-layer functionalities in two different levels~\cite{PatilEHT2019_2}. Figure~\ref{fig:FigOverview} depicts the MLD architecture, representing both MAC sub-layer levels. 

First, there is the upper MAC (U-MAC), which is a common part of the MAC sub-layer for all the interfaces. In the U-MAC, we find that link agnostic operations take place. We refer, for instance, to sequence number assignation, and MAC service data units (MSDUs) aggregation/de-aggregation. In this context, it is important to point out that the sequence number assignation must be performed at the U-MAC, since packets belonging to the same traffic flow can be fragmented and transmitted over different links. Such approach, then, eases the packet reordering at the receiver side. Additionally, common management functions for all links, such as setup, association and authentication take placed in this layer. 

Below the U-MAC, we find the low MAC (L-MAC). This lower level, which is independent for each interface, is in charge of link specific functionalities like the channel access. In this context, we find that having individual L-MAC instances allow interfaces to keep their own channel parameters if needed. Inherently, this implementation also grants each interface to keep track of their own enhanced distributed channel access (EDCA) queues (one for each access category) to hold the traffic until its transmission. Other functionalities in the L-MAC layer are the management and control frame generation, as well as the MAC header creation and validation, when transmitting and receiving respectively~\cite{LevyEHT2020}.

The motivation behind this two-tier architecture is to permit MLO-capable devices to move traffic from one link to another, being totally transparent to upper-layers. Hence, load balancing techniques may be useful to minimize the spectrum usage inefficiency of current standardized multi-band approaches, in which per client transmissions are only performed either in one band or another, by leveraging the use of all the available resources. However, such architecture entails a more complex design, requiring not only to design new methods to perform traffic to link allocation, if not also to rethink low-level aspects regarding how channel contention and packet transmissions are done in presence of multiple links. 

\subsection{Transmission modes}\label{sub:trans_modes}

The TGbe defines two different transmission modes for MLDs. First, the asynchronous transmission mode allows a MLD to transmit frames asynchronously on multiple links. Under this mode, each interface keeps its own channel access parameters with an independent behavior respect the others. Also, it allows the STR capability, enabling concurrent uplink (UL) and downlink (DL) communications, as depicted Figure~\ref{fig:FigOverview}. Ideally, it is suggested that the asynchronous mode should be selected as the default operational scheme by all 802.11be compliant nodes, since it provides a higher throughput performance~\cite{PatilEHT2019}. However, such operation must be followed by a power save mechanism, specially for handheld devices, as the power consumption may be significantly high due to having multiple asynchronous interfaces operating at the same time. Additionally, the asynchronous operation is constrained to the in-device coexistence (IDC) interference. That is, the power leakage between interfaces may prevent a frame reception on one interface, during and ongoing transmission on the other interface, as a result of not having enough separation between operating bands/channels (e.g., two channels in the 5~GHz band).

To avoid the IDC issue, the TGbe defines the synchronous mode, which relies on synchronized frame transmissions across the available links. Devices operating under a synchronous mode are referred to as constrained MLDs, or non-STR MLDs, since they are not allowed to transmit through an idle interface at the same time they are receiving through another. To perform synchronization, the end-time alignment or the defer transmission mechanisms~\cite{draft11be} may be implemented. While the former relies on ending transmissions on different channels at the same time, the latter defers the transmission of a link that has finished its backoff, until the end of the same counter in other links. With that, APs or stations are prevented to perform an STR operation, avoiding IDC problems, but at the cost of a lower throughput, if compared to the asynchronous scheme. In regards of channel access, it can be performed either following a single primary channel (SPC) or a multiple primary channel (MPC) methodology. Basically, the SPC performs contention on a unique channel, whereas in the MPC contention is performed on all channels. While applying a MPC scheme offers nodes higher chances to win contention and transmit frames, SPC allows to reduce power consumption as non-primary channel interfaces' may remain under a doze state. Figure~\ref{fig:FigOverview} shows the synchronization scheme with MPC channel access, considering the end-time alignment mechanism. Either using the SPC or the MPC method, if an interface wins contention in its channel, the others are checked during a PCF inter-frame space (PIFS) time to see if they can be aggregated, performing a transmission opportunity (TXOP) aggregation. At last, it is worth mention that MLD-capable APs may change its transmission modes (e.g., asynchronous to synchronous, and vice versa) at any time, as depicted in~Figure~\ref{fig:FigOverview}.

\subsection{Management}

\subsubsection{The Multi-Link element}\label{Mgmt:MLE}

\begin{figure}[t]
    \centering
    \includegraphics[width=\columnwidth]{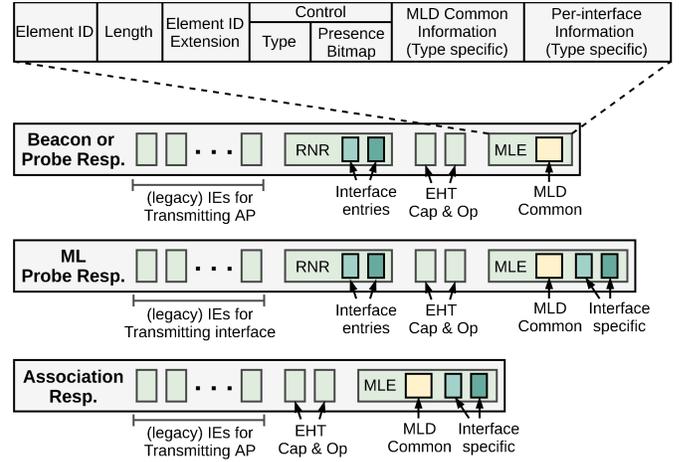}
    \caption{Multi-link element and management frames}
    \label{fig:FigMLE}
\end{figure}

The information elements (IEs) included in the different management frames allow devices to exchange their capabilities and operational parameters. With such purpose, the 802.11be defines the multi-link element (MLE). As shown in Figure~\ref{fig:FigMLE}, the MLE has been designed as a common element to the different management actions (e.g., discovery and setup). To achieve such implementation, the MLE introduces a type sub-field within the control field, that maps each operation to an specific value~\cite{ChitrakarEHT2020}. Hence, this information field is type-dependent, with its attributes announced by a presence bitmap. Such distinctive functionality provides a flexible structure to carry type specific information, while avoiding frame bloating and minimizing its overhead. 

The current 802.11be revision defines two MLE types. First, there is the basic type, which is intended to be used for beacon frames. In such type, the MLE carries only the information that is common to all interfaces. We refer, for instance, to the MLD MAC address, the set of enabled links, or the STR capability. Second, there is the multi-link request/response type, which is expected to be used during the multi-link setup. In this type, the MLE includes, apart from the common information, the complete information of those interfaces different from the advertising one, through an individual and independent field. Any parameter not advertised in the field of a given interface is considered to have the same value as the advertising one. For instance, some advertised parameters are the channel allocation (e.g., the primary channel and bandwidth), and the number of available spatial streams. Although there are currently only two defined types, further extensions of the MLE types may be aggregated. For instance, proposals are exploring to announce buffered traffic information by means of reporting a traffic indication map (TIM), or indicating changes in regards to the mapping between TID values and links. 

\subsubsection{Discovery and Setup}

The 802.11be discovery mechanism reuses the same principles already defined in the 802.11 standard. That is, stations can gather information of nearby APs by performing the discovery process based on either a passive or active scanning. However, the introduction of MLDs makes necessary to make some updates. As explained in Section~\ref{Mgmt:MLE}, beacons and probing frames only carry partial information at the multi-link level (i.e., U-MAC related-information). Such implementation, however, may take stations more time to perform the discovery process, as they should scan all the interfaces of the MLD before doing the multi-link setup. To avoid such a situation, 802.11be reuses the already defined Reduced Neighbor Report (RNR) element to announce some basic information about the different interfaces of the same AP MLD. Note that such information will belong only to the interfaces not sending the beacon frame. With that, stations can directly probe an AP~MLD requesting its complete set of capabilities, parameters and operation elements of their other interfaces. To perform such probing, they must use the multi-link request/response MLE type. Although this approach may seem inefficient, since devices need to send an extra multi-link request/response, it turns out to be the opposite as it saves energy by not requiring the non-AP MLD to enable multiple radios (i.e., scan other bands/channels of the AP MLD). Also, this approach allows to reduce the air-time occupancy of management frames, as well as, the time required by the station to pass from the discovery process to the multi-link setup. Figure~\ref{fig:FigMLE} shows the frames described.

In regards of the setup process, 802.11be will reuse the current association request/response frames by adding the extra MLE. Then, through the MLE, AP~MLDs and STA~MLDs will negotiate and establish their subsequent operation scheme by exchanging their capabilities. Besides, the multi-link setup process is proposed to be performed only on a single link in order to reduce overhead. It is worth mentioning that, the set of enabled links for each STA~MLD is determined by measuring link qualities at all interfaces. That is, those receiving a quality value above the clear channel assessment (CCA) threshold are set as enabled, while disabled otherwise. As users may keep themselves mobile, any link listed as disabled may be added afterwards by requesting a re-setup. Analogously, the re-setup process reuses the already defined re-association request/response frames. Figure~\ref{fig:FigMLE} shows the association response frame with the MLE. 

\subsubsection{Link management}\label{subsec:link}

On current multi-band APs exist the main limitation that MSDUs belonging to different TIDs are not able to be sent over multiple links. It is important to recall that TIDs were created as traffic identifiers to classify different traffic types according to their QoS needs, which establish different user priorities. Past amendments used these user priorities to provide differentiation and prioritization through EDCA, by classifying each data packet into an access category, and so, associating each one to a specific MAC transmission queue with its own MAC parameters. However, even with the use of both 2.4~GHz and 5~GHz bands in the 802.11ax amendment, all the TIDs were still tied to a single link operation. With the introduction of MLO, the IEEE 802.11be promotes such a change. By default, it is suggested for AP~MLDs to map all the TIDs to all links, implying that stations would be able to retrieve any type of traffic through any link. However, such condition may not be necessarily static, as APs may perform a dynamic TID transfer, allowing them to seamlessly move a TID from one link to another. In this context, a link management technique can be performed as a load balancing mechanism to avoid excessive levels of link congestion without performing any client steering. Indeed, this feature opens up new research opportunities in the area of load~balancing.

\subsection{Power save}\label{sub:power}

\begin{figure}[t]
  \centering
  \subfloat[Multi-link TIM]{\includegraphics[width=\columnwidth]{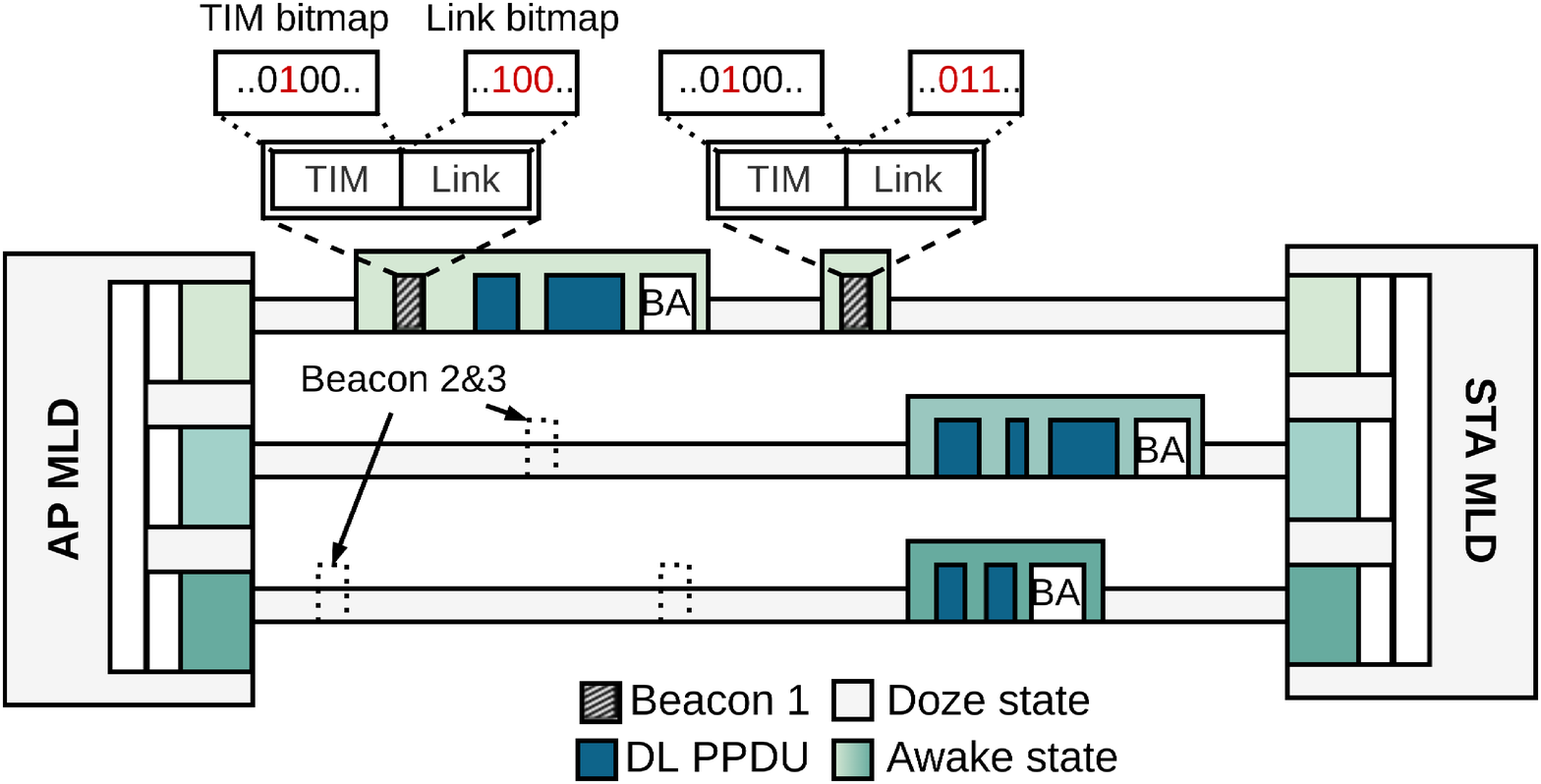}\label{subfig:TIM}}%
 \vspace{0.3cm}
  \subfloat[Multi-link TWT]{\includegraphics[width=\columnwidth]{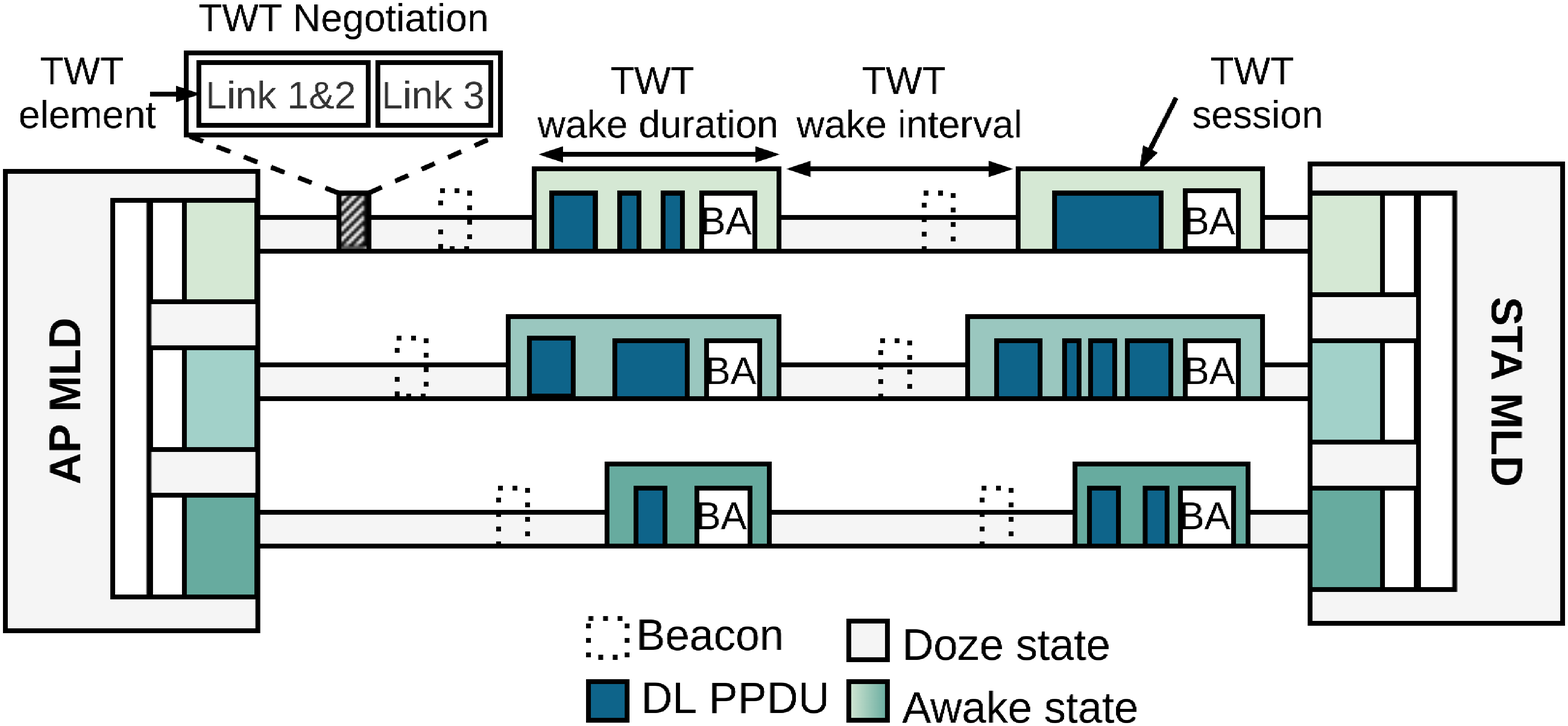}\label{subfig:TWT}}%
  \caption{Power save mechanisms}
  \label{fig:figures_power}
\end{figure}

Since the Internet connectivity nowadays is mainly performed through handheld devices, power-related consumption issues must be carefully considered. To address such issue, TGbe have suggested to adopt and adapt the use of the traffic indication map (TIM) and the target wake time (TWT).

\subsubsection{Traffic indication Map}

The TIM mechanism is used to notify stations that its serving AP has buffered data ready to be delivered to them. Thus, the AP includes the TIM element into beacons to broadcast periodically this information. In the TIM element, APs include a bitmap formed by 2007 bits, each one of them corresponding to a unique associated station. If the bit in the bitmap corresponding to a given station is 0 the station remains in a doze state, whereas if the bit is equal to 1, it goes to an awake state, being ready to retrieve the data from the AP. Although the TIM mechanism worked for single link stations, with the introduction of MLDs it has had to be revised. In order to include the information for all the multiple links that an station may be attached to, the TGbe proposed to add a link indication field following the TIM element. Within this field, a link mapping bitmap is included, where each bit indicates a designated link. Therefore, if a STA~MLD detects in the TIM element its corresponding bit set to 1, the STA~MLD further checks the link mapping, finding the specific link(s) in which the buffered traffic is mapped to~\cite{draft11be}. Figure~\ref{subfig:TIM} shows the described multi-link TIM indication mechanism. As shown, the multi-link TIM extends the classic TIM by providing an efficient functionality in which stations only need to awake determined interfaces on specific periods of time. Such a procedure, therefore, allows stations to minimize their power consumption, enlarging battery cycles.

\subsubsection{Target Wake Time}

The TWT~\cite{nurchis2019target} is a power save mechanism firstly included in the 802.11ah amendment, and further developed under the 802.11ax amendment. This mechanism relies on an initial negotiation, in which stations and APs agree in a common wake scheduling, namely session period (SP) or TWT session, where stations can send or receive data. To achieve this implementation, TWT requires from an initial negotiation phase to determine the SP parameters. To efficiently address a TWT operation under the MLO framework, TGbe suggests to perform TWT agreements (i.e., negotiation phase) for the different enabled links through a single link. To do so, STA~MLDs include in the TWT request different TWT elements, corresponding each one to a certain link that is identified through a bitmap. Such identification is needed as the links may have different TWT parameters such as wake up time, wake interval or minimum wake duration. On the contrary, if the same parameters apply for all links, only one TWT element is needed. Figure~\ref{subfig:TWT} shows the described multi-link TWT mechanism. As well as in TIM, under TWT, stations move from awake and doze states when necessary, allowing to reduce their power consumption. Although the adoption of the TWT may have different performance implications, there are no works related to such issue at the time of this article being published. Indeed, their assessment is out of scope for this paper, but an interesting topic to be addressed in future~works.

\begin{figure}[b]
    \centering
    \includegraphics[width=\columnwidth]{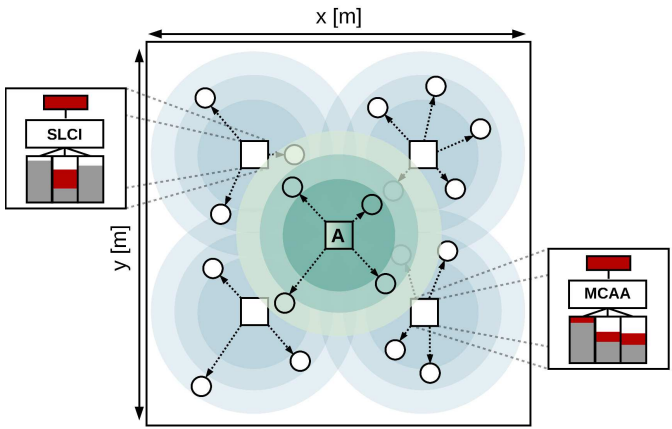}
    \caption{Scenario representation. The high, medium and low shaded areas represent the operation range for the 6~GHz, 5~GHz and 2.4~GHz bands, respectively.}
    \label{fig:scenario}
\end{figure}

\section{Traffic management}\label{trafic_mgmt}

The existence of APs and stations with multiple interfaces makes traffic flow allocation a challenging part for the MLO. In this regard, the 802.11be U-MAC implementation should rely on a traffic manager to distribute the buffered data to a certain L-MAC, as the efficient use of the interfaces will play a critical role in terms of network performance. To that end, we introduced a set of policies in~\cite{lopez2021ieee}. First, \textit{Multi Link Same Load to All interfaces} (MLSA) allocates traffic equally to all interfaces. However, it was demonstrated that such operation is highly inefficient, as the channel occupancy is not taken into account when driving decisions. Then, we also introduced the \textit{Single Link Less Congested Interface} (SLCI) and the \textit{Multi-link Congestion-Aware load balancing at flow Arrivals} (MCAA) policies. They rely on link occupancy measurements to allocate traffic either to a single or multiple bands, resulting in significant performance gains as expected. Hence, we showcased that traffic decisions must take into account the instantaneous occupancy of the channel, as well as the own traffic load. However, the non-adaptive implementation of the proposed mechanisms may not be efficient for long-lasting flows, as links may change its occupancy very rapidly. In this context, we set for further study the adoption of a dynamic strategy that not only takes into account the instantaneous channel occupancy of each interface when a flow becomes active, but tracks them continuously, so the traffic can be reallocated dynamically when changes~happen.

Although the previous strategies considered all TIDs to be mapped into the multiple interfaces, the MLO opens up the possibility to perform a link-based traffic separation through the TID-to-link mapping functionality. That is, different TIDs may be mapped to different links, in order to minimize, for instance, access delays for time-sensitive traffic. Besides, such feature may be complemented by the fact that nodes' spatial distribution may create different contention-free links, specially in the 5~GHz and 6~GHz bands, as a result of favorable radio propagation conditions. Therefore, traffic with higher QoS requirements can be exclusively exchanged through those contention-free links, as long as they exist.

To showcase the benefits and drawbacks in the application of a TID-to-link mapping strategy, in this paper we introduce a new traffic allocation policy that distinguishes between traffic flows of different types. That is, traffic corresponding to data flows will be allocated to different links than the video flows. We will refer to this policy as \textit{Video and Data Separation} (VDS), and it will allocate data flows to the 2.4~GHz or 5~GHz band, whereas video flows will be allocated at the 6~GHz band. Following the results from~\cite{lopez2021ieee}, VDS will not distribute data flows across multiple interfaces, but it will allocate the whole traffic to a single interface (i.e., either interface at 2.4~GHz or 5~GHz band, selecting always the emptiest one).

\begin{table}
   \small 
   \centering 
   \caption{Evaluation setup}
   \resizebox{\linewidth}{!}{
    \begin{tabular}{ll} 
       \toprule
       \textbf{Parameter} & \textbf{Description}\\ 
       \midrule
       Carrier frequency & 2.437 GHz/5.230 GHz/6.295 GHz\\
       Channel bandwidth & 20 MHz/40 MHz/80 MHz\\
       AP/STA TX power & 20/15 dBm\\
       CCA threshold & -82 dBm\\
       AP/STA noise figure & 7 dB\\
       \makecell[l]{Single user \\ spatial streams} & 2\\
       MPDU payload size & 1500 bytes\\
       Path loss & Same as \cite{lopez2021ieee}\\
       Avg. data duration & T$_{\text{ON}}=$ 3 s\\
       \makecell[l]{Avg. data \\ interarrival time}  & T$_{\text{OFF}}=$ 1 s\\
       \makecell[l]{Min. contention \\ window} & 15\\
       Packet error rate & 10\%\\
       Simulation time (1 simulation) & 120 s\\
       Number of deployments & N$_{\text{D}}=$ 500 \\
       \bottomrule
       \end{tabular}}\label{subtbl:the-table}
\end{table}


\section{Performance evaluation}\label{sec:perf_eval}

This section aims to conduct a flow-level performance analysis of a link-based traffic allocation strategy under the MLO framework. Simulations are done using the CSMA/CA abstraction presented in \cite{lopez2020concurrent}. We evaluate N$_{\text{D}}=$ 500 random generated deployments, all of them with 5~BSSs as depicted in Figure~\ref{fig:scenario}. Each BSS consists of one AP and $M$~stations placed around it. In every deployment, we will place the $\text{BSS}_\text{A}$ at the center, and the other $4$~BSSs  distributed uniformly at random over a 20x20 $\text{m}^{2}$ area. 

Unless stated otherwise, we consider that all MLD AP/STAs are configured with 3 wireless interfaces that operate at a different frequency band (i.e., 2.4~GHz, 5~GHz and 6~GHz). All stations are inside the coverage area of its AP for at least the 2.4~GHz interface, as shown in Figure~\ref{fig:scenario}. For evaluation purposes, APs' interfaces corresponding to the same frequency band are configured with the same radio channel. Except for $\text{AP}_\text{A}$, which will be set either with the SLCI, MCAA, VDS or MSLA, the rest of the APs will implement either the SLCI or MCAA policies, selected with the same probability.

Only DL traffic is considered. Upon creation, stations request either a video or a data traffic flow. The two type of flows are alive during the entire simulation time, but their activity follows an ON/OFF Markovian process. The ON and OFF periods are exponentially distributed with mean duration T$_{\text{ON}}$ and T$_{\text{OFF}}$. For each individual video (data) flow in the ON period, the corresponding AP has to deliver $\ell_S$~($\ell_E$)~Mbps. Table~\ref{subtbl:the-table} details the complete set of parameters used.

\begin{figure}[t]
  \centering
  \includegraphics[width=.95\linewidth]{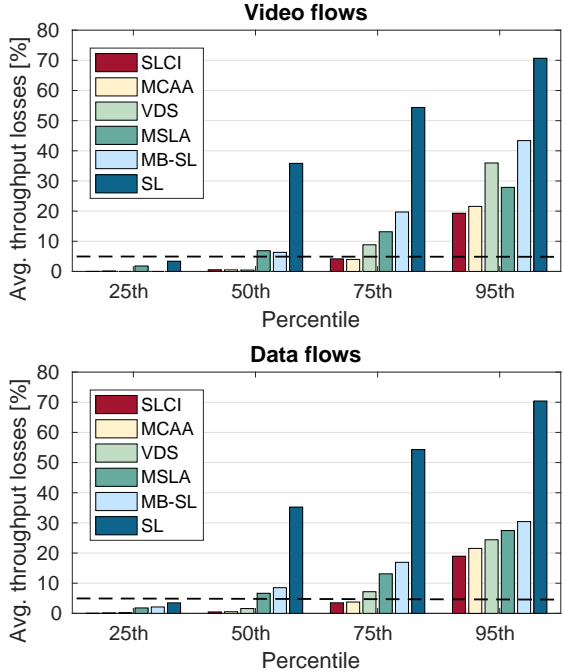}%
    \caption{Avg. throughput losses for video and data flows. Black dashed line corresponds to the 5\% losses threshold.}
  \label{fig:dratio}
\end{figure}

Figure~\ref{fig:dratio} shows different percentiles of the average throughput losses suffered by video and data flows for each policy through the different N$_{\text{D}}$. As observed, the VDS policy is able to keep the average throughput losses under a 5\% value for both video and data flows only in the 50\% of the evaluated scenarios. In fact, it is noticeable that for video flows, the 5\% worst case raises up to throughput losses nearly 40\%, performing even worse than the MSLA policy. Such results reveal a critical drawback of the VDS policy: the traffic separation in VDS may suffer from severe performance problems in conditions with high neighboring BSSs overlap, or high traffic scenarios. On the other hand, the SLCI and MCAA congestion-aware policies are able to overcome such negative issues in 75\% of the scenarios due to their ability to balance the traffic load between the active links.

At last, also in Figure~\ref{fig:dratio}, we provide a comparison between MLO-based, and both legacy multi-band single link (MB-SL) and legacy single link (SL) deployments. Through the MB-SL, we clearly observe the advantage of adding more bands to the system, as the stations can be spread across them, reducing also their congestion levels. Although the MB-SL performance is better when compared to SL, it barely keeps the throughput losses below an acceptable 5\% value for both flow types in only 25\% of the scenarios. Compared to the legacy approaches, MLO is shown to be able to perform better in all the evaluated scenarios independently. In fact, we observe that with SL and MB-SL only the 25 \% of the considered scenarios achieve average throughput losses below 5 \%, which is increased up to the 75 \% with either SLCI and MCAA. Those results, prove that the MLO framework will be a relevant new feature to WiFi, enabling currently unsuitable scenarios with SL and MB-SL solutions.


\section{Open issues and challenges}\label{open}

Although the MLO represents a promising functionality to be implemented in next generation WLANs, the concurrent use of multiple interfaces brings new challenges to face off. In this context, we point out some open issues that require further research:

\begin{itemize}
    \item \textit{non-STR and legacy blindness}. This issue relates to the fact that non-STR and legacy STAs may cause different collision scenarios, as a consequence of their constrained operation. First, non-STR STAs may be unable to detect an intra-BSS transmission (either DL or UL) in one of its available links, because of performing a transmission on another. Therefore, a collision may occur if non-STR STAs attempt to transmit over that link already in use. This issue has been already tackled in \cite{naribole2020simultaneous} by allowing AP~MLDs to inform non-STR stations about the channel state in other links in use by the AP~MLD to prevent such a situation. On the other hand, similarly, legacy devices may not know if a transmission is taking place in others links, since they only operate in a single one. Hence, some indication, as the proposed in the non-STR case, is needed to inform legacy nodes of the activities happening in the other links.
    \item \textit{Spectrum inefficiency}. Conservative approaches to avoid the IDC interference or collisions can lead to an inefficient use of the spectrum, because of suspending the backoff procedure in one link, if medium access is granted in another one. In this regard, an opportunistic backoff mechanism to maximize the spectrum utilization of non-STR nodes is proposed in \cite{naribole2020simultaneous}, so transmission attempts can be resumed only when the channel state guarantees a collision with not happen.
    \item \textit{Channel access fairness}. Since MLO allows to perform TXOP aggregation over different links, nodes with single link availability may experience starvation due to their higher difficulties to access the channel. Therefore, in presence of legacy stations the usage of link aggregation techniques should be limited or restricted, in order to minimize unfair situations.
    \item \textit{Load balancing}. Although this has been the main topic of this paper, further research is required to fully understand which is the best strategy to balance the traffic in MLO WLANs. For instance, it is important to consider also how MLO can be used for uplink traffic, as it may require a completely different approach than its downlink counterpart. In this aspect, load balancing strategies can benefit from the use of machine learning solutions to predict future traffic and network dynamics.
\end{itemize}


\section{Concluding remarks}

This article has overviewed the EHT MLO framework with the objective to provide a clear and concise understanding of this upcoming disruptive WiFi functionality. The MLO framework will allow next generation of APs and stations to perform concurrent transmissions by using their multiple wireless interfaces in a coordinated way, and therefore, opening the door to both improve the network performance and achieve a more efficient use of the spectrum resources. However, further research need to be done to fully understand all new features enabled by MLO. Apart from the traffic allocation, we identified other open issues that must be tackled such as the spectrum inefficiency and the channel access fairness.

\bibliographystyle{unsrt}
\bibliography{main}

\begin{thebibliography}{10}

\bibitem{draft11be}
{IEEE P802.11be/D1.0 Draft Standard for Information technology—
  Telecommunications and information exchange between systems Local and
  metropolitan area networks— Specific requirements. Part 11: Wireless LAN
  Medium Access Control (MAC) and Physical Layer (PHY) Specifications.
  Amendment 8: Enhancements for extremely high throughput (EHT)}, May 2021.

\bibitem{lopez2019ieee}
David L{\'o}pez-P{\'e}rez, Adrian Garcia-Rodriguez, Lorenzo Galati-Giordano,
  Mika Kasslin, and Klaus Doppler.
\newblock {IEEE 802.11 be Extremely High Throughput: The Next Generation of
  Wi-Fi Technology beyond 802.11 ax}.
\newblock {\em IEEE Communications Magazine}, 57(9):113--119, 2019.

\bibitem{naik2021can}
Gaurang Naik, Dennis Ogbe, and Jung-Min~Jerry Park.
\newblock {Can Wi-Fi 7 Support Real-Time Applications? On the Impact of Multi
  Link Aggregation on Latency}.
\newblock In {\em ICC 2021-IEEE International Conference on Communications},
  pages 1--6. IEEE, 2021.

\bibitem{lacalle2021analysis}
Guillermo Lacalle, I{\~n}aki Val, Oscar Seijo, Mikel Mendicute, Dave
  Cavalcanti, and Javier Perez-Ramirez.
\newblock {Analysis of Latency and Reliability Improvement with Multi-Link
  Operation over 802.11}.
\newblock In {\em 2021 IEEE 19th International Conference on Industrial
  Informatics (INDIN)}, pages 1--7. IEEE, 2021.

\bibitem{carrascosa2021experimental}
Marc Carrascosa, Giovanni Geraci, Edward Knightly, and Boris Bellalta.
\newblock {An Experimental Study of Latency for IEEE 802.11 be Multi-link
  Operation}.
\newblock {\em arXiv preprint arXiv:2111.09281}, 2021.

\bibitem{naribole2020simul}
Sharan Naribole, Srinivas Kandala, Wook~Bong Lee, and Ashok Ranganath.
\newblock {Simultaneous Multi-Channel Downlink Operation in Next Generation
  WLANs}.
\newblock In {\em GLOBECOM 2020-2020 IEEE Global Communications Conference},
  pages 1--7. IEEE, 2020.

\bibitem{naribole2020simultaneous}
Sharan Naribole, Wook~Bong Lee, Srinivas Kandala, and Ashok Ranganath.
\newblock {Simultaneous Transmit-Receive Multi-Channel Operation in Next
  Generation WLANs}.
\newblock In {\em 2020 IEEE Wireless Communications and Networking Conference
  (WCNC)}, pages 1--8. IEEE, 2020.

\bibitem{yang2019ap}
Mao Yang, Bo~Li, Zhongjiang Yan, and Yuan Yan.
\newblock {AP Coordination and Full-duplex enabled Multi-band Operation for the
  Next Generation WLAN: IEEE 802.11 be (EHT)}.
\newblock In {\em 2019 11th International Conference on Wireless Communications
  and Signal Processing (WCSP)}, pages 1--7. IEEE, 2019.

\bibitem{PatilEHT2019_2}
Abhishek Patil, George Cherian, Alfred Asterjadhi, and Duncan Ho.
\newblock {Multi-Link Operation: Design Discussion}.
\newblock IEEE~802.11~Documents. 2019. [Online]. Available:
  \url{https://mentor.ieee.org/802.11/documents?is_dcn=823&is_group=00be}.

\bibitem{LevyEHT2020}
Joseph Levy and Xiaofei Wang.
\newblock {802.11be Architecture/Association Discussion}.
\newblock IEEE~802.11~Documents. 2020. [Online]. Available:
  \url{https://mentor.ieee.org/802.11/documents?is_dcn=1122&is_group=00be}.

\bibitem{PatilEHT2019}
Abhishek Patil, George Cherian, Alfred Asterjadhi, and Duncan Ho.
\newblock {Multi-Link Aggregation - Gain Analysis}.
\newblock IEEE~802.11~Documents. 2019. [Online]. Available:
  \url{https://mentor.ieee.org/802.11/documents?is_dcn=764&is_group=00be&is_year=2019}.

\bibitem{ChitrakarEHT2020}
Rojan Chitrakar, Rajat Pushkarna, Yanyi Ding, and Yoshio Urabe.
\newblock {Multi-link element format}.
\newblock IEEE~802.11~Documents. 2020. [Online]. Available:
  \url{https://mentor.ieee.org/802.11/documents?is_dcn=772&is_group=00be}.

\bibitem{nurchis2019target}
Maddalena Nurchis and Boris Bellalta.
\newblock {Target wake time: Scheduled access in IEEE 802.11 ax WLANs}.
\newblock {\em IEEE Wireless Communications}, 26(2):142--150, 2019.

\bibitem{lopez2021ieee}
{\'A}lvaro L{\'o}pez-Ravent{\'o}s and Boris Bellalta.
\newblock {IEEE 802.11 be Multi-Link Operation: When the Best Could Be to Use
  Only a Single Interface}.
\newblock In {\em 2021 19th Mediterranean Communication and Computer Networking
  Conference (MedComNet)}. IEEE, 2021.

\bibitem{lopez2020concurrent}
{\'A}lvaro L{\'o}pez-Ravent{\'o}s and Boris Bellalta.
\newblock {Concurrent Decentralized Channel Allocation and Access Point
  Selection using Multi-Armed Bandits in multi BSS WLANs}.
\newblock {\em Computer Networks}, 180, 2020.

\end{thebibliography}

\begin{IEEEbiography}[{\includegraphics[width=1in,height=1.25in,clip,keepaspectratio]{{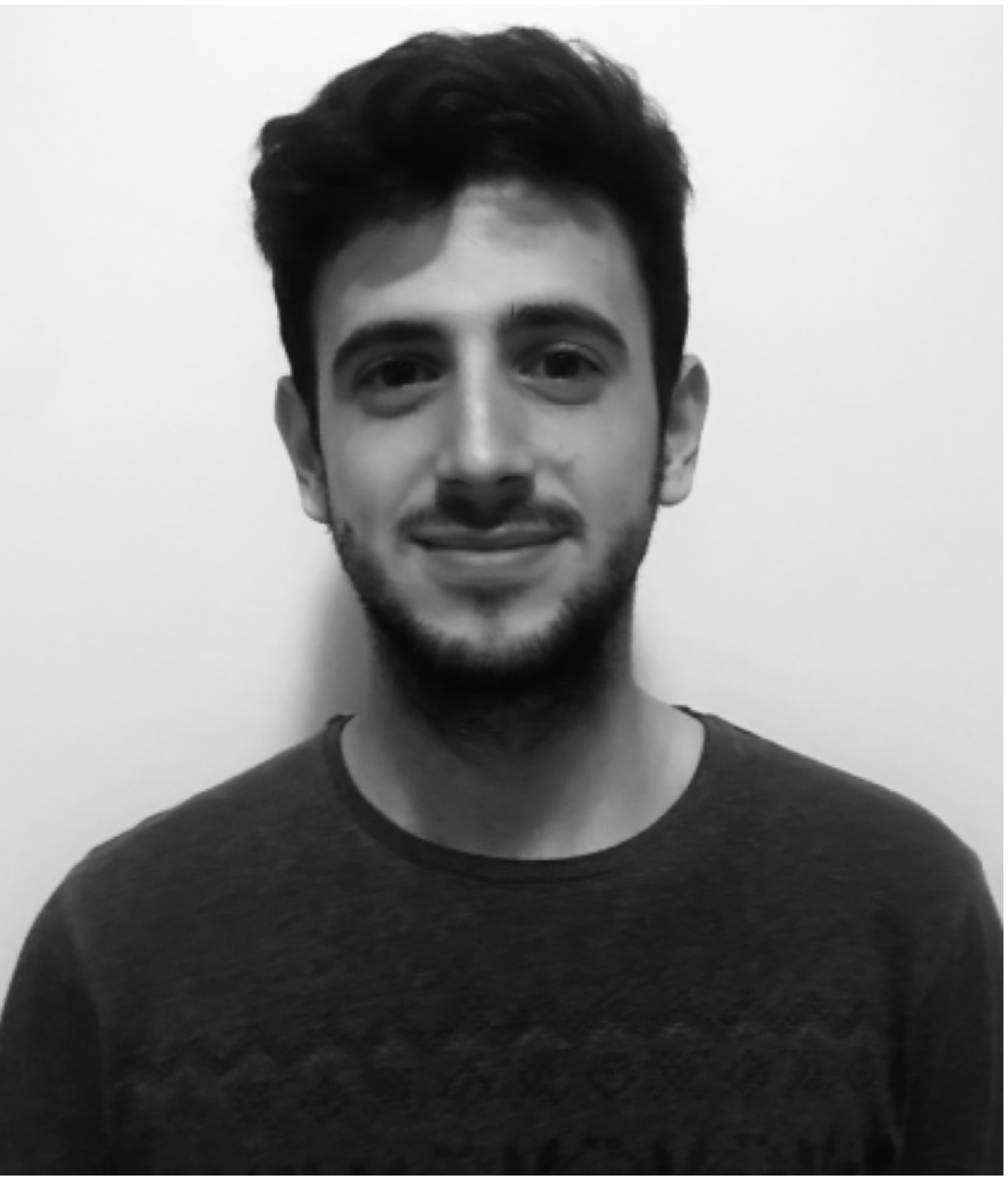}}}]{Álvaro~López-Raventós} received his B.Sc. degree in telecommunications systems engineering from Universitat Politècnica de Catalunya (UPC), and his M.Sc. in wireless communications systems from Universitat Pompeu Fabra (UPF) in 2016 and 2017, respectively. Currently, he is carrying out a Ph.D. in the Wireless Networking (WN) research group at UPF. His research interests cover programmable wireless networks, with emphasis on machine learning techniques for network optimization and performance management.\end{IEEEbiography}

\begin{IEEEbiography}[{\includegraphics[width=1in,height=1.25in,clip,keepaspectratio]{{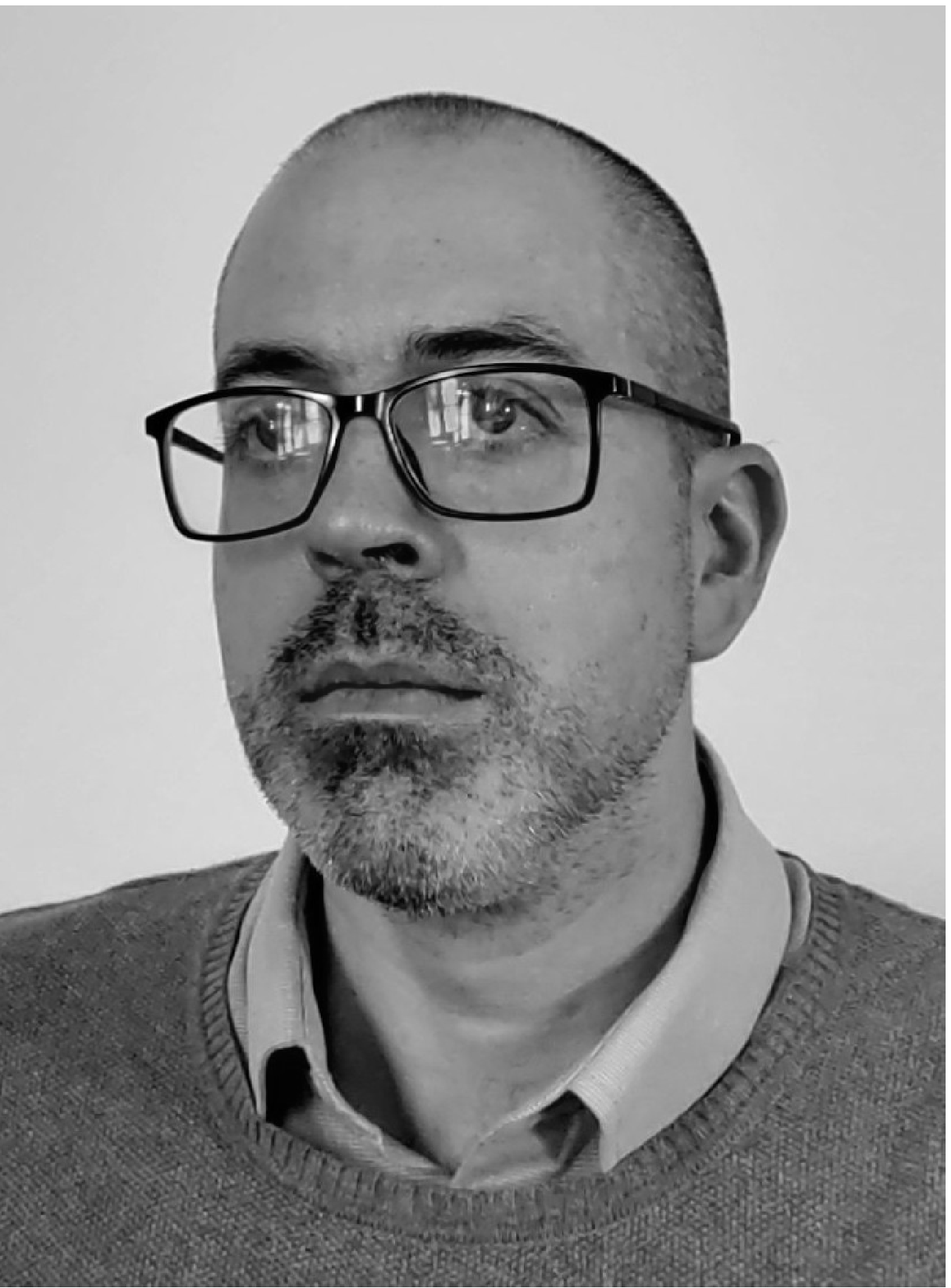}}}]{Boris Bellalta} is an associate professor in the Department of Information and Communication Technologies (DTIC) at the Universitat Pompeu Fabra (UPF). His research interests are in the area of wireless networks, with an emphasis on the design and performance evaluation of new architectures and protocols.\end{IEEEbiography}

\end{document}